\documentclass[english,3p]{elsarticle}
\usepackage{lmodern}
\usepackage[T1]{fontenc}
\usepackage[latin9]{inputenc}
\usepackage{xcolor}
\usepackage{pdfcolmk}
\usepackage{amsmath}
\usepackage{amssymb}
\usepackage{graphicx}
\PassOptionsToPackage{normalem}{ulem}
\usepackage{ulem}
\usepackage[unicode=true,
 bookmarks=true,bookmarksnumbered=false,bookmarksopen=false,
 breaklinks=false,pdfborder={0 0 1},backref=false,colorlinks=true]
 {hyperref}
\hypersetup{
 pdfborderstyle=}

\makeatletter

\usepackage{babel}
\usepackage{array}
\usepackage{bm}
\usepackage{multirow}
\usepackage{braket}
\usepackage{enumitem}
\usepackage{float}
\usepackage{extarrows}
\usepackage[numbers]{natbib}

\providecommand{\tabularnewline}{\\}

\providecommand{\doi}[1]{%
	\begingroup
	\let\bibinfo\@secondoftwo
	\urlstyle{rm}%
	\href{http://dx.doi.org/#1}{%
		\discretionary{}{}{}%
		\nolinkurl{#1}%
	}%
	\endgroup
}

\usepackage{xcolor}
\usepackage{textcomp}
\usepackage{hyphenat}
\usepackage{makecell}

\usepackage{listings}
\usepackage{xcolor}
\definecolor{darkred}{rgb}{0.5 0 0}
\definecolor{darkgreen}{rgb}{0.5 .5 0}
\definecolor{darkblue}{rgb}{0 0 .5}

\usepackage{algorithm}
\usepackage[noend]{algpseudocode}

\newtheorem{prop}{Proposition}
\newtheorem{defi}{Definition}

\makeatother

\begin{document}

\title{\textsf{MagneticKP}: A package for quickly constructing \textbf{\textit{k}}$\cdot$\textbf{\textit{p}} models of magnetic and non-magnetic crystals}

\author[buct,bit1,bit2]{Zeying Zhang}

\author[bit1,bit2,sutd]{Zhi-Ming Yu\corref{cor}}
\cortext[cor]{Corresponding author}
\ead{zhiming_yu@bit.edu.cn}

\author[bit1,bit2]{Gui-Bin Liu}
\author[buct]{Zhenye Li}

\author[sutd]{Shengyuan A. Yang\corref{cor}}
\ead{shengyuan_yang@sutd.edu.sg}


\author[bit1,bit2]{Yugui Yao}
	




\address[buct]{College of Mathematics and Physics, Beijing University of Chemical
Technology, Beijing 100029, China}

\address[bit1]{Centre for Quantum Physics, Key Laboratory of Advanced Optoelectronic Quantum Architecture and Measurement (MOE), School of Physics, Beijing Institute of Technology, Beijing, 100081, China}

\address[bit2]{Beijing Key Lab of Nanophotonics \& Ultrafine Optoelectronic Systems, School of Physics, Beijing Institute of Technology, Beijing, 100081, China}

\address[sutd]{Research Laboratory for Quantum Materials, Singapore University of Technology and Design, Singapore 487372, Singapore}

\date{\today}
\begin{abstract}

We propose an efficient algorithm to construct $\boldsymbol{k}\cdot \boldsymbol{p}$ effective Hamiltonians, which is much faster than the previously proposed algorithms. This algorithm is implemented in \textsf{MagneticKP} package. The package applies to both single-valued (spinless) and double-valued (spinful) cases, and it works for both magnetic and nonmagnetic systems.
By interfacing with \textsf{SpaceGroupIrep} or \textsf{MSGCorep} packages, it can directly output the $\boldsymbol{k}\cdot \boldsymbol{p}$ Hamiltonian around arbitrary momentum and expanded to arbitrary order in $k$.
\\



\textbf{Program summary}

Program title: \textsf{MagneticKP}


Licensing provisions: GNU General Public Licence 3.0

Programming language: Mathematica

External routines/libraries used: \textsf{SpaceGroupIrep} (Optional), \textsf{MSGCorep} (Optional)

Developer's repository link: https://github.com/zhangzeyingvv/MagneticKP

Nature of problem: Construct $\boldsymbol{k}\cdot\boldsymbol{p}$ Hamiltonian for arbitrary magnetic space group

Solution method: Linear algebra, iterative algorithm to solve common null space of operators


\end{abstract}
\maketitle


\section{Introduction}


$\boldsymbol{k}\cdot\boldsymbol{p}$ modelling is widely used in the research of condensed matter physics. Such models describe the local band structure around certain momentum $\bm K$ in the Brillouin zone and take the form of a Taylor expansion in powers of $\bm k$, with $\bm k$ being the derivation from $\bm K$. The famous early examples include the Kohn-Luttinger model and the Kane model for studying semiconductor materials \cite{luttinger_motion_1955,kane_band_1957}. In the past twenty years, with the development in two-dimensional materials and topological materials, $\boldsymbol{k}\cdot\boldsymbol{p}$ modelling has become a standard tool for studying their properties. In these materials, the physical responses are mostly determined by the electronic states around a few band extremal or degeneracy points, so $\boldsymbol{k}\cdot\boldsymbol{p}$ models are most suitable for their description. For example, many exotic properties of graphene can be understood from its 2D Dirac model obtained using  $\boldsymbol{k}\cdot\boldsymbol{p}$ method \cite{castro_neto_electronic_2009}. $\boldsymbol{k}\cdot\boldsymbol{p}$ models have also been constructed to study other 2D semiconductors, such as transition-metal dichalcogenides and monolayers of group-IV or group-V elements \cite{liu_quantum_2011,liu_low-energy_2011,xiao_coupled_2012,lu_multiple_2016}, to capture the band inversion topology such as in HgTe quantum wells \cite{bernevig_quantum_2006}, and to describe nodal states in topological semimetals, such as Weyl/Dirac points \cite{wan_topological_2011,wang_dirac_2012,yang_classification_2014,young_dirac_2015}, triple points \cite{weng_topological_2016,zhu_triple_2016}, and various nodal loops/surfaces \cite{PhysRevLett.113.046401, weng_topological_2015, zhao_unified_2016, bzdusek_robust_2017, wu_nodal_2018}.


In practice, a $\boldsymbol{k}\cdot\boldsymbol{p}$ model is usually constructed from symmetry constraint. The input information include the symmetry group of at the expansion point $\bm K$ and the symmetry information of the target band states at $\bm K$. Depending on the needs, the output model is expanded to a specified cutoff power of $k$.
At present, there already exist a few packages, including \textsf{kdotp-symmetry}\cite{gresch_identifying_2018}, \textsf{Qsymm} \cite{varjas_qsymm_2018}, \textsf{kdotp-generator} (based on \textsf{kdotp-symmetry}) \cite{jiang_kp_2021} and \textsf{Model-Hamiltonian} \cite{zhan_programmable_2021}, which can construct $\boldsymbol{k}\cdot\boldsymbol{p}$  Hamiltonians.
All these  packages are
written in Python and use a similar algorithm, namely, the direct-product decomposition algorithm (DDA).
In the DDA approach, each symmetry constraint is transformed to a set of linear equations, and one solves the null space of these equations by the standard linear algebra method. After going through all symmetry constraints, one obtains a collection of null spaces. The output model Hamiltonian is obtained by calculating the intersection of all the null spaces using the standard Zassenhaus algorithm \cite{luks_algorithms_1997}, such that it satisfies all the symmetry constraints.

In this work, we propose an improved algorithm, which
has been implemented in our \textsf{MagneticKP} package (written in Wolfram language). We term this algorithm as the iterative simplification algorithm (ISA). We show that compared with the DDA, ISA reduces the time complexity of constructing $\boldsymbol{k}\cdot \boldsymbol{p}$ Hamiltonians.
The improvement increases with the symmetry group size, the model dimension, and the cutoff power in $k$.
The main difference  lies in the method to obtain the intersection of a collection of null spaces. As mentioned above, DDA uses the direct Gaussian elimination method, which is quite time consuming. Instead, ISA adopts an iterative method, such that the problem size is reduced at each step in obtaining the common null spaces of two operators. Besides the improvement in algorithm, the usage of Wolfram language in the \textsf{MagneticKP} package also helps to enhance the speed, since
its handling of analytic calculation is more efficient than Python. The application and the validity of our algorithm and package have been demonstrated in many of our previous works \cite{zhang_coexistence_2017, yu_encyclopedia_2021,liu_systematic_2022,zhang_encyclopedia_2022}.


This paper is organized as follows: In Sec.~2, we give a detailed description of ISA and compare it with DDA. In Sec.~3, we introduce the capabilities of \textsf{MagneticKP} package, including the installation and running of \textsf{MagneticKP}. In Sec.~4, we present a simple example.
Finally, a conclusion is given in Sec.~5.

\section{Algorithm}
\label{sec:alg}

To construct a $\boldsymbol{k}\cdot \boldsymbol{p}$ Hamiltonian, we need to first specify the expansion point $\bm K$ and the
basis states at $\bm K$. The form of the $\boldsymbol{k}\cdot \boldsymbol{p}$ Hamiltonian is constrained by the symmetry elements of the little co-group $G$ at $\bm K$. Consider a symmetry $Q\in G$. Its constraint on the Hamiltonian $H$ is given by
\begin{equation}\label{ham}
H(\boldsymbol{k})=
\begin{cases}
	D(Q)H(R^{-1}\boldsymbol{k})D^{-1}(Q)& \text{if } Q = \{R|t\}\\
	D(Q)H^*(-R^{-1}\boldsymbol{k})D^{-1}(Q)& \text{if } Q = \{R|t\}{\cal T}\\
\end{cases}
\end{equation}
where the relation depends on whether $Q$ involves the time reversal operation $\mathcal{T}$, $D(Q)$ is the matrix
(co)representation matrix of $Q$ (not necessarily irreducible) in the basis states. The target result is a Hamiltonian that satisfies symmetry constraints by all the $Q$'s in $G$ and meanwhile includes all the allowed terms. In the calculation, one does not need to go through all the $Q$'s. Only the generators of the magnetic little co-group at $\bm K$ are needed.

%

\subsection{Problem formulation}
Now, we formulate the above problem into a form that can be handled numerically. Suppose we take $N$ basis states at $\bm K$, and we demand a model expanded to $P$-th power in $k$. We may first decompose the $\boldsymbol{k}\cdot \boldsymbol{p}$ Hamiltonian $H$ as a sum
\begin{equation}
  H(\bm k)=\sum_{m=0}^P \mathcal{H}_m(\bm k),
\end{equation}
where each $\mathcal{H}_m$ includes terms that of $m$-th power in $k$. According to (\ref{ham}), the symmetry transforms $\bm k$ in a linear way, so each individual $\mathcal{H}_m$ would satisfy the symmetry constraint in (\ref{ham}). Therefore, we are allowed to consider each $\mathcal{H}_m$ separately. 

Note that $H$ and $\mathcal{H}_m$'s are $N\times N$ complex Hermitian matrices. It is known that $N\times N$ complex Hermitian matrices form a vector space over $\mathbb{R}$, which has a dimension of $N^2$. We can choose $N^2$ basis for this vector space, and label them as $M_{\mu}$ with $\mu=1,\cdots, N^2$. For example, for $N=2$, the four basis may be chosen as the identity matrix and the three Pauli matrices; for $N=3$, one may choose the identity and the eight Gell-Mann matrices, and so on.

After choosing the basis $M_\mu$, we can express the Hamiltonian $\mathcal{H}_m(\bm k)$ in the following form
\begin{equation}
\begin{aligned}
	\mathcal{H}_m(\boldsymbol{k})
	&=\sum_{\ell=1,..,L;\ \mu=1,..,N^2} c^{\ell \mu} p_\ell(\boldsymbol{k})M_\mu. \\
\end{aligned}
\label{eq:hk}
\end{equation}
{Here, $p_\ell(\boldsymbol{k})\in\{k_x^a k_y^b k_z^c|a+b+c= m; a,b,c\geq0\}$  is a product of the $k$ vector components with a total power of $m$. There are totally 
$L=\frac{1}{2} (m+1) (m+2)$ such products. We label these products by $\ell$, which runs from $1$ to $L$. 
The expansion coefficients $c^{\ell\mu}\in \mathbb{R}$ are what we want to find after imposing the symmetry constraints. 

First, consider the first line in (\ref{ham}), i.e., for the case when $Q=\{R|t\}$ not involving $\mathcal{T}$.  Note that the $D$ matrix does not depend on $\bm k$. Then the right hand of Eq.~(\ref{ham}) (for $\mathcal{H}_m$) can be expressed as
\begin{equation}
\begin{aligned}
	D(Q)\mathcal{H}_m(R^{-1}\boldsymbol{k})D^{-1}(Q)
	=\sum_{\ell,n=1,\dots,L;\mu,\nu=1,..,N^2}c^{n\mu}p_\ell(\boldsymbol{k})F^\ell_{\ n}(Q)M_\nu J_{\ \mu}^{\nu}(Q).\\
\end{aligned}
\end{equation}
Here, $F^\ell_{\ n}(Q)$ is an $L\times L$ constant matrix satisfying $p_n(R^{-1}\boldsymbol{k})=\sum_{\ell}p_\ell(\boldsymbol{k})F^\ell_{\ n}$; and $J_{\ \mu}^{\nu}(Q)$ is an $N^2\times N^2$ matrix satisfying $DM_\mu D^{-1}=\sum_{\nu} M_\nu J_{\ \mu}^{\nu}$.
Then, the symmetry condition in (\ref{ham}) can be re-written in the following form
\begin{equation}
	\begin{aligned}
	\mathbf{0}=&\mathcal{H}_m(\boldsymbol{k})-D(Q)\mathcal{H}_m(R^{-1}\boldsymbol{k})D^{-1}(Q)\\
=&\sum_{\ell,n=1,\dots,L;\mu,\nu=1,..,N^2}p_\ell(\bm k)M_\nu \Big[\delta^{\ell}_{\ n} \delta_{\ \mu}^\nu-F^\ell_{\ n}(Q) J^\nu_{\ \mu}(Q)\Big]
c^{n\mu}\\
	=& \mathbf{b}\cdot \mathbf{S}(Q)\cdot \mathbf{c}\\
	\end{aligned}
\label{eq:minus}
\end{equation}
where $\mathbf{b}\equiv(p_1 M_1, p_1 M_2\cdots,p_L M_{N^2})$ is a $1\times LN^2$ row vector, $\mathbf{c}\equiv(c^{11},\cdots, c^{LN^2})^T$ is a
$LN^2\times 1$ column vector, and
\begin{equation}
  \mathbf{S}(Q)\equiv \delta^{\ell}_{\ n} \delta_{\ \mu}^\nu-F^\ell_{\ n}(Q) J^\nu_{\ \mu}(Q)
\end{equation}
is an $L N^2 \times LN^2$ matrix. 
Since $\mathbf{b}$ is a vector of linearly independent $k$-products, the condition in (\ref{eq:minus}) is equivalent to
\begin{equation}
  \mathbf{S}(Q)\ \mathbf{c}=\mathbf{0}. 
\end{equation}
Thus, $\mathbf{c}\in \mathrm{ker}\ \mathbf{S}(Q)$, so the condition reduces to finding the null space or the kernel of  $\mathbf{S}(Q)$.

As for the second line in Eq.~(\ref{ham}), i.e., for $Q = \{R|t\}{\cal T}$, one can see that we only need to slightly modify the definition of $\mathbf{S}(Q)$ as
\begin{equation}
  \mathbf{S}(Q)\equiv \delta^{\ell}_{\ n} \delta_{\ \mu}^\nu-\tilde{F}^\ell_{\ n}(Q) \tilde{J}_{\ \mu}^{\nu}(Q),
\end{equation}
where $\tilde{F}^\ell_{\ n}(Q)$ is defined from the relation $p_n(-R^{-1}\boldsymbol{k})=\sum_{\ell}p_\ell(\boldsymbol{k})\tilde{F}^\ell_{\ n}$; and $\tilde{J}^\nu_{\ \mu}(Q)$ satisfies $D M_\mu^* D^{-1}=\sum_{\nu} M_\nu \tilde{J}_{\ \mu}^{\nu}$.


Typically, the group $G$ has multiple generators $Q_1, Q_2, \cdots$. Each generator $Q_i$ gives a $\mathbf{S}(Q_i)$ for which we solve its null space $\mathrm{ker}\ \mathbf{S}(Q_i)$. The final solution is their common subspace $\mathfrak{S}=\bigcap_i \mathrm{ker}\ \mathbf{S}(Q_i)$. In practice, we need to solve out a basis set $\{\mathbf{u}_1,\cdots, \mathbf{u}_r\}$ for $\mathfrak{S}$, where $r=\text{dim}\mathfrak{S}$, such that the coefficient $c^{\ell\mu}$ in Eq.~(\ref{eq:hk}) is expressed as
\begin{equation}
  c^{\ell \mu}=\sum_{i=1}^r a^i [\mathbf{u}_i]^{\ell\mu}
\end{equation}
with $r$ real coefficients $a^i$ serving as the model parameters for $\mathcal{H}_m$. 



\subsection{Iterative simplification  algorithm}
The direct way to obtain the common null space $\mathfrak{S}$ is by the Gaussian elimination method. Here, we propose an iterative numerical method. To this end, we need to first introduce a definition, a proposition and a short proof. 

{

\begin{defi}
	$S$ is a linear mapping from $\mathbb{C}^m$ to $\mathbb{C}^n$. The matrix for  $S$  in the standard bases is denoted as $\mathbf{S}$, which is of size $n\times m$. Let $\{\alpha_1,\alpha_2,\cdots, \alpha_r\}$ be any basis set for $\mathrm{ker}\, \mathbf{S}$,  i.e., $\mathrm{ker}\, \mathbf{S}=\mathrm{Span}\{\alpha_1, \alpha_2, ..., \alpha_r\}$, $r=\mathrm{dim}\,\mathrm{ker}\, \mathbf{S}$. Then, we define a $m\times r$ matrix $\mathcal{K}(\mathbf{S})$ associated with $\mathbf{S}$ by
%
%
	\begin{equation*}
		\mathcal{K}(\mathbf{S})=(\alpha_1, \alpha_2, ..., \alpha_r).
	\end{equation*}
When treating each column of $\mathcal{K}(\mathbf{S})$ as a vector in $\mathbb{C}^m$, we can write
\begin{equation*}
  \mathrm{ker}\, \mathbf{S}=\mathrm{Span}\{\mathcal{K}(\mathbf{S})\}.
\end{equation*}
\end{defi}

\begin{prop}
	\label{prop}
	Consider two linear mappings $A:\mathbb{C}^m\rightarrow\mathbb{C}^n$ and $B:\mathbb{C}^m\rightarrow\mathbb{C}^p$. The  matrices for  ${A}$ and ${B}$ in standard bases are $\mathbf{A}$ and $\mathbf{B}$, which are of size $n\times m$ and $p\times m$, respectively. Then, we have
	\begin{equation*}
		 \mathrm{ker}\, \mathbf{A}\bigcap \mathrm{ker}\, \mathbf{B}=\mathrm{Span}\big\{\mathcal{K}(\mathbf{A}) \cdot \mathcal{K}(\mathbf{B}\cdot \mathcal{K}(\mathbf{A}))\big\}.
	\end{equation*}	
\end{prop}
{\it Proof:}
The intersection space of $\ker A $ and $\ker B$ is equal to the kernel of linear mapping restricted to the space $\ker A$, i.e. $\ker A \cap \ker B =\ker ( B|_{\ker A})$. Now, take a set of bases of $\ker \mathbf{A}:$ $\{\alpha_1, \alpha_2, ..., \alpha_r\}$ and let $\mathcal{K}(\mathbf{A})=(\alpha_1, \alpha_2, ..., \alpha_r)$ being a $m\times r$ matrix. The linear mapping $B|_{\ker A}:$ $\ker A\rightarrow \mathbb{C}^n$ in  the basis of $\{\alpha_1, \alpha_2, ..., \alpha_r\}$  is represented by the $p\times r$ matrix $\mathbf{B}\cdot \mathcal{K}(\mathbf{A})$. Then the $r\times \ell$ matrix $\mathcal{K}(\mathbf{B}\cdot \mathcal{K}(\mathbf{A}))$ gives the basis set of $\ker A\bigcap \ker B$ expressed in the basis of $\{\alpha_1, \alpha_2, ..., \alpha_r\}$, where $\ell=\mathrm{dim} (\ker A \cap \ker B)$. The multiplication of $\mathcal{K}(\mathbf{A})$ from the left converts them back to the standard bases, which generates the desired result.
\\


As we have discussed in the last subsection, the target is to find a basis set for the common null subspace $\mathfrak{S}=\bigcap_{i=1}^s \mathrm{ker}\ \mathbf{S}(Q_i)$, where $s$ is the number of generators of $G$. Based on the above proposition, we can obtain it in the following iterative way. Let $\mathcal{U}_1=\mathcal{K}(\mathbf{S}(Q_1))$, and for $1\leq i\leq s-1$,
\begin{equation}
\mathcal{U}_{i+1}=\mathcal{U}_i \cdot \mathcal{K}(\mathbf{S}(Q_{i+1})\cdot \mathcal{U}_i)
\end{equation}
Then the final matrix $\mathcal{U}_s=(\mathbf{u}_1,\cdots, \mathbf{u}_r)$ contains the desired basis set for $\mathfrak{S}$.

The pseudo code for obtaining  $\mathcal{U}_s$ is shown in Algorithm~\ref{alg:is}.
\begin{algorithm}[H]
	\caption{Iterative calculation of $\mathcal{U}_s$}\label{alg:is}
	\begin{algorithmic}
		\Procedure{$\mathcal{U}$}{{\{$\mathbf{S}(Q_1),...,\mathbf{S}(Q_s)$\}}}
		\State$\mathcal{U}\gets \mathcal{K}(\mathbf{S}(Q_1))$
		\For {$\mathbf{S}$ in \{$\mathbf{S}(Q_2),...,\mathbf{S}(Q_s)$\}}
		\If {$\mathcal{U}=\varnothing$} \Return $\varnothing$
		\EndIf
		\State $\mathcal{U}\gets  \mathcal{U} \cdot \mathcal{K}(\mathbf{S}\cdot \mathcal{U})$
		\EndFor
		\Return $\mathcal{U}$
		\EndProcedure
	\end{algorithmic}
\end{algorithm}


}

\subsection{Comparison of ISA to DDA}

DDA differs from ISA in the way to obtain the common null space basis set, i.e., the $\mathcal{U}_s$ above.
In DDA, after constructing $\mathbf{S}(Q_i)$ $(i=1,\cdots,s)$, one needs to first solve the null space matrix $\mathcal{K}(\mathbf{S}(Q_i))$ for each $i$ separately. Then, $\mathcal{U}_s$ is obtained by combining all the
$\mathcal{K}(\mathbf{S}(Q_i))$'s into one big matrix and do Gaussian elimination to find the common basis set \cite{gresch_identifying_2018}. Hence, it is a two-step process. 

In comparison, in ISA, $\mathcal{U}_s$ is obtained in an iterative way. The dimension of space, i.e., the size of matrix, decreases in each iteration step, which greatly reduces the computational cost. The processes of the two algorithms are illustrated in Fig.~\ref{fig:compare}.

{

\begin{figure}[t]
	\includegraphics[width=\columnwidth]{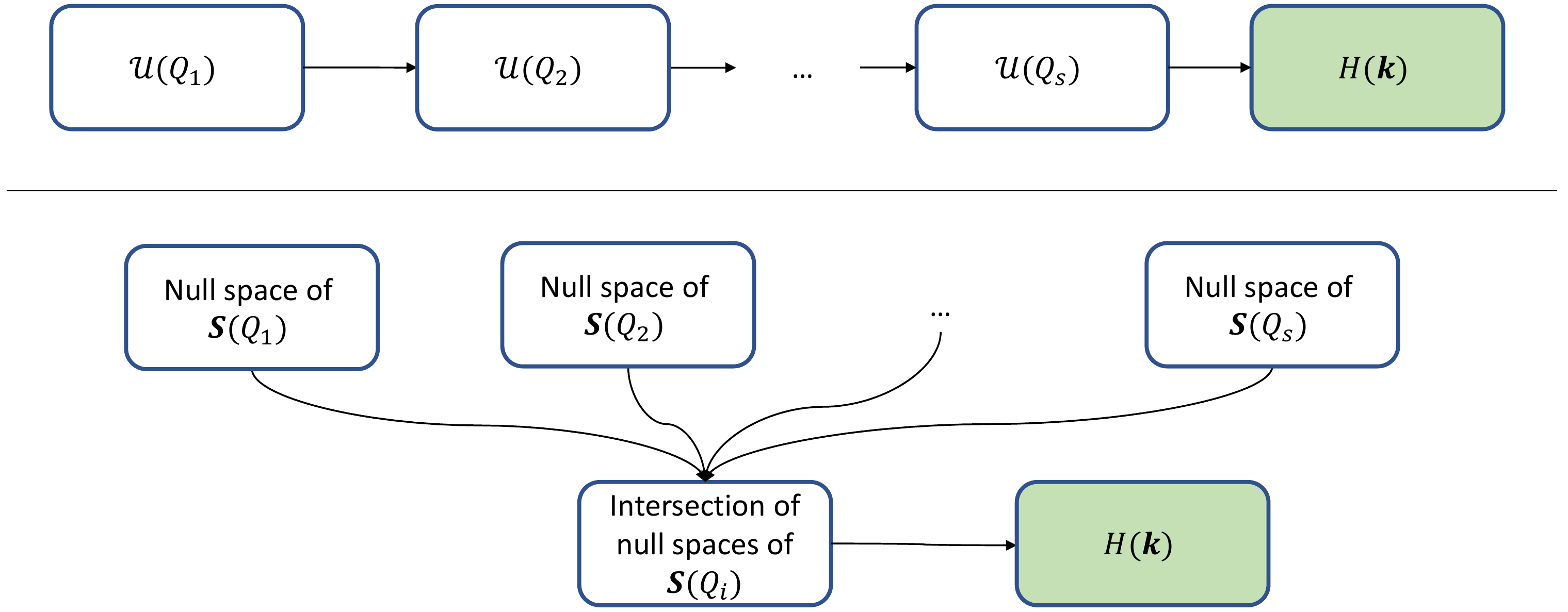}
	\caption{Schematic comparison of the two algorithms. Upper panel: ISA. Lower panel: DDA.
	}
	\label{fig:compare}
\end{figure}

Here, we give an estimation of the time complexity of the two algorithms.
The complexity for calculating the null space of a $a\times b$ $(a \geq b)$ matrix is approximately $O(ab^2)$ \cite{hogben_handbook_2006} (note that this is a very rough estimate, because the method for \lstinline|NullSpace| in Wolfram language is automatically chosen by \lstinline|"CofactorExpansion"|, \lstinline|"DivisionFreeRowReduction"| and \lstinline|"OneStepRowReduction"|, thus the complexity of \lstinline|NullSpace|
also depends on the specific form of the matrix). The time complexity for finding $\mathcal{U}_s$ in DDA is about $O(s(LN^2)^3+ (\sum_{i=1}^s d_i)\times (LN^2)^2)$, where $d_i=\mathrm{dim} \ker \mathbf{S}(Q_i)$. The first term is for calculating the null spaces for the $s$ matrices $\mathbf{S}(Q_i)$, and the second term is for calculating their intersection. In comparison, for ISA,  the   time complexity is approximately
$
O((LN^2)^3+LN^2\sum_{n=2}^{s}(LN^2-\sum_{i=1}^{n}r_i)^2)
$, where $r_i$ is the number of columns of $\mathcal{U}_i$. In practice, we find the first term dominates, so ISA complexity is roughly $O((LN^2)^3)$, which is at least $s$ times faster than DDA.

}

To test the computational efficiency, we construct several $\boldsymbol{k}\cdot \boldsymbol{p} $ Hamiltonians using three different ways: (1) ISA implemented in \textsf{MagneticKP} (written in Wolfram language), (2) DDA implemented in \textsf{MagneticKP} (written in Wolfram language), and (3) DDA implemented in \textsf{kdotp-symmetry} (written in Python language). The test results as shown in Table~\ref{tab:mats}. 
One can see that ISA implemented in \textsf{MagneticKP} has the best performance.
The time cost difference between approaches (1) and (3) becomes more and more pronounced with the cutoff power and basis size. One also notes that approach (2) is also much better than (3), which demonstrates that 
for the current task involving analytic calculations, Wolfram language is more efficient than Python.

\begin{table}[tb]
	\centering
	\caption{Comparison of time costs for three different approaches.  The column labeled "MSG" gives the magnetic space group number. "corep" gives the $\Gamma$ label of the co-representation (basis size), "dim" is the dimension of the co-representation, 	"$k$-order" is the cutoff power of the $\boldsymbol{k}\cdot\boldsymbol{p}$ model. The three approaches are:  (1) ISA implemented in \textsf{MagneticKP},  (2) DDA implemented in \textsf{MagneticKP}, and (3)  DDA implemented in \textsf{kdotp-symmetry}. The values are in unit of second. All tests are run on intel i7-10870H CPU@2.20GHz with 32GB RAM.\\}
	\begin{tabular}{@{\extracolsep{\fill}}*{1}{ccccccc}@{}}
		\hline
		\hline
		MSG & corep & dim &  $\boldsymbol{k}$-order	& ISA (\textsf{MagneticKP}) & DDA (\textsf{MagneticKP}) & DDA (\textsf{kdotp-symmetry}) \tabularnewline
		\hline
		226.123  &$L_4L_4$	&4 & 2   & 0.43  &  0.59   & 5.76 \tabularnewline
		& 	&  & 4   & 1.18 &  3.20   & 137.99 \tabularnewline
		& 	&  & 6   & 3.30 &  12.12   & > 2 hours \tabularnewline
		& 	&  & 8   & 8.80 &  36.43   & > 2 hours \tabularnewline
		\hline
		218.82  &$R_4R_5$	&6 & 2   & 4.48  &  5.97   & 41.58 \tabularnewline
		& 	&  & 4   & 10.22 &  22.80   & 490.33 \tabularnewline
		& 	&  & 6   & 28.87 &  86.55   & > 2 hours \tabularnewline
		& 	&  & 8   & 87.87 &  282.83  & > 2 hours \tabularnewline
		\hline
	\end{tabular}
	\label{tab:mats}
\end{table}

\section{Capability of \textsf{MagneticKP}}

\label{sec:Capabilities}

\subsection{Installation}
The steps of installing \textsf{MagneticKP} is exactly the same as installing \textsf{MagneticTB} \cite{zhang_magnetictb_2021}.
One just needs to unzip the "MagneticKP-main.zip" file and copy
the MagneticKP directory to any  directory in \lstinline!$Path!. e.g.,
copy to \lstinline!FileNameJoin[{$UserBaseDirectory, "Applications"}]!.
Then, one can start to use the package after running \lstinline!Needs["MagneticKP`"]!.
The version of Mathematica should be $\ge$ v11.3.

\subsection{Running}

\subsubsection{Core module}

The core part of \textsf{MagneticKP} package is the function \lstinline!kpHam! which computes the $\boldsymbol{k}\cdot\boldsymbol{p}$ Hamiltonian. The format of
this function is
\begin{lstlisting}[backgroundcolor={\color{yellow!5!white}},mathescape=true]
kpHam[korder, input, "Method"->"IterativeSimplification" or "DirectProductDecomposition"]
\end{lstlisting}
Here, \lstinline!korder! can be both an integer or a list of integers that specifies the cutoff power in $k$ for the  $\boldsymbol{k}\cdot\boldsymbol{p}$ Hamiltonian to be calculated.
When \lstinline!korder! is a list such as $\{n,m\}$,  the function will output two Hamiltonians of the cutoff power of $n$ and $m$, respectively.
\lstinline!input!  has the format of an \lstinline!Association! in Mathematica. It contains the input information for constructing the Hamiltonian.
There are three necessary inputs, the
rotation part of $Q$, the (co)representation matrix of $Q$, and whether $Q$ is an unitary or an anti-unitary operator.
The format of \lstinline!input! is
\begin{lstlisting}[backgroundcolor={\color{yellow!5!white}},mathescape=true]
input = <|
"Unitary" -><|$Q_1$ -> {$D(Q_1)$, $R_1\boldsymbol{k}$},...|>,
"Anitunitary" -><|$Q_2$ -> {$D(Q_2)$, -$R_2\boldsymbol{k}$},...|>
|>
\end{lstlisting}
Notice that the role of \lstinline|Keys| of \lstinline|input["Unitary"]| or \lstinline|input["Anitunitary"]| is to make the input  clearer, \textsf{MagneticKP} will respectively  read the \lstinline|Values| of \lstinline|input["Unitary"]| and \lstinline|input["Anitunitary"]|  to do the calculation. $R\boldsymbol{k}$ can be in either Cartesian or primitive coordinates.

The default method in \lstinline!kpHam! is  ISA. Users can explicitly specify a method by  putting \lstinline!"Method"->"IterativeSimplification"! or
\lstinline!"Method"->"DirectProductDecomposition"! in \lstinline!kpHam!.
After the above parameters are set appropriately, one can run \lstinline!kpHam! to obtain the $\boldsymbol{k}\cdot\boldsymbol{p}$ Hamiltonian. The output of \lstinline!kpHam! is also an \lstinline!Association!.
The format of the output  is [see Sec.~\ref{sec:example} for a concrete example]
\begin{lstlisting}[backgroundcolor={\color{yellow!5!white}},mathescape=true]
<|"ham" -> expression of $\boldsymbol{k}\cdot\boldsymbol{p}$ Hamiltonian,  "korder" -> order of Hamiltonian,
"dim" -> dimension of Hamiltonian, "NumberOfParameters" -> number of parameters|>
\end{lstlisting}

\subsubsection{IO module}

The input to \lstinline!kpHam! contains the matrix $D(Q_i)$. Here, we introduce how to get its expression.
In general, for $Q_i\in G$, one can use the projective representation method to get the irreducible representation $\Delta(Q_i)$. The reality of $\Delta(Q_i)$ can be determined by Herring's rule \cite{herring_effect_1937} and $D(Q_i)$ can be easily constructed from $\Delta(Q_i)$ and the reality of $\Delta(Q_i)$ \cite{bradley_mathematical_2009}.
More direct method is to obtain $D(Q_i)$ from standard reference books \cite{bradley_magnetic_1968,bradley_mathematical_2009,lax_symmetry_2001,dresselhaus_group_2008}, or Bilbao crystallographic server \cite{aroyo_bilbao_2006}, or many software packages \cite{bosma_magma_1997, iraola_irrep_2020, matsugatani_qeirreps_2021, noauthor_gap_2021, liu_spacegroupirep_2021}. Here, we provide a function \lstinline|interfaceRep| to interface with packages \textsf{SpaceGroupIrep} \cite{liu_spacegroupirep_2021} and \textsf{MSGCorep} \cite{liu_MSGCorep_2021}.
\textsf{MSGCorep} package is our home-made package and  will be made public soon. The format of \lstinline|interfaceRep| is
\begin{lstlisting}[backgroundcolor={\color{yellow!5!white}},mathescape=true]
interfaceRep[MSGNO, k, reps, "CartesianCoordinates" -> True or False, "CalculateGenerators" -> True or False]
\end{lstlisting}
where \lstinline|MSGNO| can be either space group number (one integer) or BNS magnetic space group number (a list containing two integers). When \lstinline|MSGNO| is an integer number (list), \textsf{SpaceGroupIrep} (\textsf{MSGCorep}) package must be loaded.   \lstinline|k| can be given in the form of the coordinate of $\boldsymbol{K}$ or the symbol of the $\boldsymbol{K}$ point (if it is a high-symmetry point).
\lstinline|reps| is an integer or a list of integers, which represents  the serial number of irreducible (co)representations in \lstinline|showLGIrepTab|(\lstinline|showMLGCorep|). When \lstinline|reps| is a list, \textsf{MagneticKP} will automatically
calculate the direct sum of (co)representations. \lstinline|"CartesianCoordinates"| (default value is \lstinline|True|) tells \textsf{MagneticKP} whether to convert the operations into Cartesian coordinates. Finally, since \textsf{SpaceGroupIrep} (\textsf{MSGCorep}) will show all the
symmetry operations in the (magnetic) little group, to save the computing resources we develop a greedy algorithm to find the generators of a group \cite{cormen_introduction_2001}. The pseudo code is shown in Algorithm~\ref{alg:greedy}.
\begin{algorithm}
	\caption{Use greedy algorithm to find the generators of a group}\label{alg:greedy}
	\begin{algorithmic}
		\Procedure{getGenerator}{InputGroup}
		\State Generator $\gets$ $\varnothing$
		\State group $\gets$  \{identity element\}
		\For {element in InputGroup}
		\State temGenerator $\gets$ \lstinline!Append[Generator,element]!
		\State temgroup $\gets$ \lstinline!GenerateGroup[temGenerator]!
		\If {group $\neq$ temgroup}
		Generator $\gets$ temGenerator; \State group $\gets$ temgroup
		\EndIf
		\If {group $=$ InputGroup} Break
		\EndIf
		\EndFor
		\Return Generator
		\EndProcedure
	\end{algorithmic}
\end{algorithm}

It should be mentioned that 
Ref.~\cite{jiang_kp_2021,tang_exhaustive_2021} only generate models for high symmetry $\boldsymbol{k}$ points. In comparison, in \textsf{MagneticKP}, with the help of \textsf{SpaceGroupIrep} (\textsf{MSGCorep}),
it can work for arbitrary $\boldsymbol{k}$ point,  for arbitrary direct sums of more than two irreducible representations, for different types of coordinates etc. Hence, it is also more general and more convenient than previous packages. 


\section{Example}
\label{sec:example}
We use the four-band nodal ring in TiB$_2$ \cite{zhang_coexistence_2017} as an example to show how to use \textsf{MagneticKP}.  TiB$_2$ is a nonmagnetic material with time-reversal symmetry and belongs to   space group 191 ($P6/mmm$) (see Fig.~\ref{fig:TiB2}(a)). The four-band nodal ring appears around  $K\ (-\frac{1}{3},\frac{2}{3},0)$ point when spin-orbit coupling effect is neglected. The generators of the little co-group at $K$ can be chosen as $\{C_{3}^+|000\}$, $\{C_{2}''|000\}$, $\{\sigma_h|000\}$ and $\{I{\cal T}|000\}$. Then $R\boldsymbol{k}$ and the single-value representation matrices of the relevant band representations ($K_5\oplus K_6$) can be written as \cite{bradley_mathematical_2009}:
\begin{equation}
\begin{aligned}
C_{3}^+:(k_x,k_y,k_z)&\rightarrow(-\frac{k_x}{2}-\frac{\sqrt{3} k_y}{2},\frac{\sqrt{3} k_x}{2}-\frac{k_y}{2},k_z),&D(C_{3}^+)=-\frac{\Gamma _{0,0}}{2}-\frac{1}{2} i \sqrt{3} \Gamma _{3,2} \\
C_{2}'':(k_x,k_y,k_z)&\rightarrow(k_x,-k_y,k_z), &D(C_{2}'')=\Gamma _{0,3} \\
\sigma_h:(k_x,k_y,k_z)&\rightarrow(k_x,k_y,-k_z), &D(\sigma_h)={\Gamma _{3,0}}\\
I{\cal T}:(k_x,k_y,k_z)&\rightarrow(k_x,k_y,k_z), &D(I{\cal T})=\Gamma _{0,0}\\
\end{aligned}
\end{equation}
where $\Gamma_{i,j}=\sigma_{i}\otimes\sigma_{j}$,  $\sigma_{0}$ is the $2\times 2$ identity matrix and $\sigma_{i} (i=1,2,3)$ are the three Pauli matrices. With these input information, one can run the following code to get the $\boldsymbol{k}\cdot \boldsymbol{p}$ Hamiltonian up to first order
\begin{lstlisting}[backgroundcolor={\color{yellow!5!white}},mathescape=true,numbers=left]
Needs["MagneticKP`"];
input=<|"Unitary" -> <|
C3 -> {-IdentityMatrix[4]/2 + I Sqrt[3] KroneckerProduct[PauliMatrix[3], PauliMatrix[2]]/2, {-kx/2 - (Sqrt[3] ky)/2, Sqrt[3] kx/2 - ky/2, kz}},
C2 -> {KroneckerProduct[PauliMatrix[0], PauliMatrix[3]], {-kx, ky, -kz}},
$\sigma$h -> {KroneckerProduct[PauliMatrix[3], PauliMatrix[0]], {kx, ky, -kz}}|>,
"Anitunitary" -> <|IT -> {IdentityMatrix[4], {kx, ky, kz}}|>|>;
MatrixForm[kpHam[1, input]["ham"]]
\end{lstlisting}
The output of the above script is:
\begin{center}
	\includegraphics[width=\linewidth]{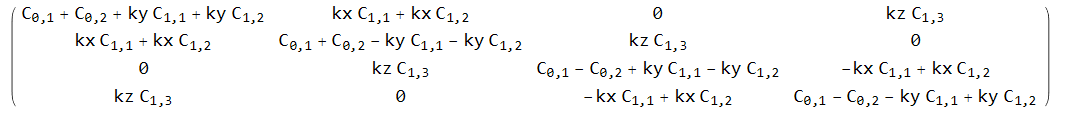}
\end{center}
Here, $C_{i,j}$ is the $j$-th real parameter of the $i$-th $\boldsymbol{k}$-order of the $\boldsymbol{k}\cdot\boldsymbol{p}$ Hamiltonian. On the $k_z = 0$ plane, the  Hamiltonian is decoupled into two $2\times2$ diagonal
blocks, which has different mirror eigenvalues ($+1$ for $K_5$ and $-1$ for $K_6$) and makes it possible to generate a nodal ring on the $k_z=0$ plane.
To fully capture  the four-band nodal ring in TiB$_2$~\cite{zhang_coexistence_2017}, one needs a $2$nd order Hamiltonian, which can be easily obtained by  changing  line 7 in the above script to
\begin{lstlisting}[backgroundcolor={\color{yellow!5!white}},mathescape=true]
MatrixForm[kpHam[2, input]["ham"]]
\end{lstlisting}	
{ The band structure of the 2nd order $\boldsymbol{k}\cdot \boldsymbol{p}$ Hamiltonian is shown in Fig.~\ref{fig:TiB2}(b), which is consistent with the result in Ref.~\cite{zhang_coexistence_2017}.}

A more direct way is to  interface with \textsf{MSGCorep}. One needs to simply write
\begin{lstlisting}[backgroundcolor={\color{yellow!5!white}},mathescape=true]
Needs["MSGCorep`"]
input = interfaceRep[{191, 234}, "K", {5, 6}];
kpHam[2, input]
\end{lstlisting}
Here, the output of \lstinline|interfaceRep| is:
\begin{center}
\includegraphics[width=\linewidth]{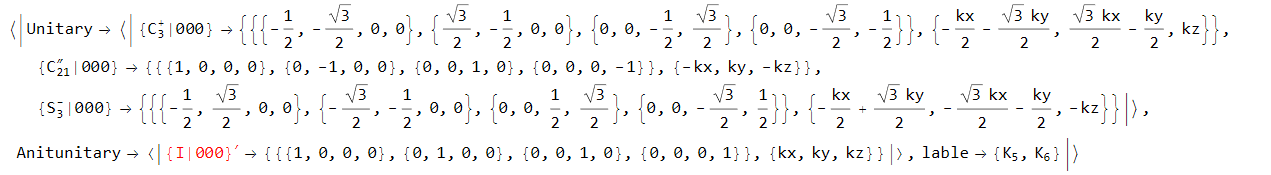}
\end{center}
This output  additionally contains the labels of (co)representations, which would make the analysis more convenient. 

\begin{figure}[t]
	\centering
	\includegraphics[width=0.8\columnwidth]{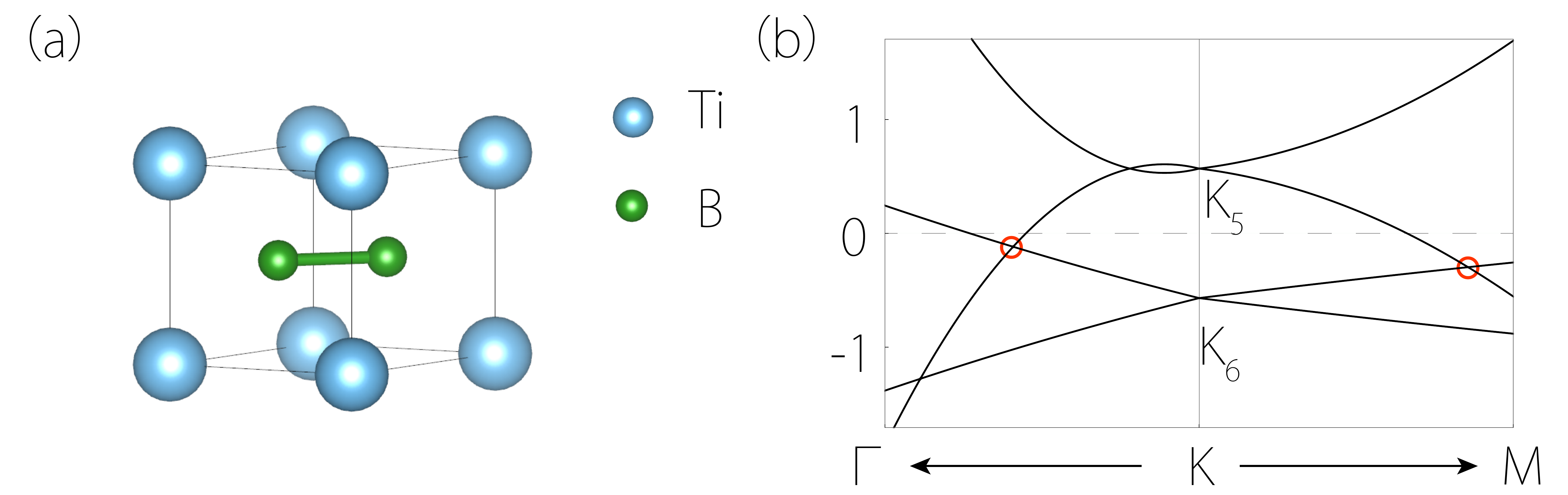}
	\caption{(a) Crystal structure of TiB$_2$. (b) Band structure of the output of \lstinline|kpHam[2, input]|. Here, we take the values $C_{0,1}= 0,C_{0,2}= 0.6,C_{2,1}= 0,C_{2,2}= 0.13,C_{2,3}= 0.,C_{2,4}= 0.12,C_{2,5}= 0,C_{2,6}= 0,C_{2,7}= 0,C_{1,1}= 0.1,C_{1,2}= 0,C_{1,3}= 0$. The two red circles indicate the crossing points on the nodal ring.
	}
	\label{fig:TiB2}
\end{figure}


\section{Conclusion}

In conclusion, we have developed  a package \textsf{MagneticKP} to
generate $\boldsymbol{k}\cdot \boldsymbol{p}$ Hamiltonian at an arbitrary momentum point. We develop the ISA approach, which is much faster than the algorithm used in previous packages. By interfacing with \textsf{SpaceGroupIrep} (\textsf{MSGCorep}),
\textsf{MagneticKP} can generate the $\boldsymbol{k}\cdot \boldsymbol{p}$ Hamiltonians for any (magnetic) space group.
The package will be a useful tool for band structure modeling and analysis.

\section*{Acknowledgments}

This work is supported by the National Key R\&D Program of China
(Grant No. 2020YFA0308800), the NSF of China (Grants  No.~12004035,
No.~12004028, No.~11734003, No.~12061131002, and No.~52161135108), the Strategic Priority Research
Program of Chinese Academy of Sciences (Grant No.~XDB30000000),  the Beijing Natural Science
Foundation (Grant No.~Z190006), and the Singapore Ministry of Education
AcRF Tier 2 (Grant No.~T2EP50220-0026).

\bibliographystyle{elsarticle-num-names}
\bibliography{MagneticKP1}

\begin{thebibliography}{44}
\expandafter\ifx\csname natexlab\endcsname\relax\def\natexlab#1{#1}\fi
\providecommand{\url}[1]{\texttt{#1}}
\providecommand{\href}[2]{#2}
\providecommand{\path}[1]{#1}
\providecommand{\DOIprefix}{doi:}
\providecommand{\ArXivprefix}{arXiv:}
\providecommand{\URLprefix}{URL: }
\providecommand{\Pubmedprefix}{pmid:}
\providecommand{\doi}[1]{\href{http://dx.doi.org/#1}{\path{#1}}}
\providecommand{\Pubmed}[1]{\href{pmid:#1}{\path{#1}}}
\providecommand{\bibinfo}[2]{#2}
\ifx\xfnm\relax \def\xfnm[#1]{\unskip,\space#1}\fi
\bibitem[{Luttinger and Kohn(1955)}]{luttinger_motion_1955}
\bibinfo{author}{J.~M. Luttinger}, \bibinfo{author}{W.~Kohn},
\newblock \bibinfo{title}{Motion of {Electrons} and {Holes} in {Perturbed}
  {Periodic} {Fields}},
\newblock \bibinfo{journal}{Physical Review} \bibinfo{volume}{97}
  (\bibinfo{year}{1955}) \bibinfo{pages}{869}.
\bibitem[{Kane(1957)}]{kane_band_1957}
\bibinfo{author}{E.~O. Kane},
\newblock \bibinfo{title}{Band structure of indium antimonide},
\newblock \bibinfo{journal}{Journal of Physics and Chemistry of Solids}
  \bibinfo{volume}{1} (\bibinfo{year}{1957}) \bibinfo{pages}{249--261}.
\bibitem[{Castro~Neto et~al.(2009)Castro~Neto, Guinea, Peres, Novoselov, and
  Geim}]{castro_neto_electronic_2009}
\bibinfo{author}{A.~H. Castro~Neto}, \bibinfo{author}{F.~Guinea},
  \bibinfo{author}{N.~M.~R. Peres}, \bibinfo{author}{K.~S. Novoselov},
  \bibinfo{author}{A.~K. Geim},
\newblock \bibinfo{title}{The electronic properties of graphene},
\newblock \bibinfo{journal}{Reviews of Modern Physics} \bibinfo{volume}{81}
  (\bibinfo{year}{2009}) \bibinfo{pages}{109--162}.
\bibitem[{Liu et~al.(2011{\natexlab{a}})Liu, Feng, and Yao}]{liu_quantum_2011}
\bibinfo{author}{C.-C. Liu}, \bibinfo{author}{W.~Feng},
  \bibinfo{author}{Y.~Yao},
\newblock \bibinfo{title}{Quantum {Spin} {Hall} {Effect} in {Silicene} and
  {Two}-{Dimensional} {Germanium}},
\newblock \bibinfo{journal}{Physical Review Letters} \bibinfo{volume}{107}
  (\bibinfo{year}{2011}{\natexlab{a}}) \bibinfo{pages}{076802}.
\bibitem[{Liu et~al.(2011{\natexlab{b}})Liu, Jiang, and
  Yao}]{liu_low-energy_2011}
\bibinfo{author}{C.-C. Liu}, \bibinfo{author}{H.~Jiang},
  \bibinfo{author}{Y.~Yao},
\newblock \bibinfo{title}{Low-energy effective {Hamiltonian} involving
  spin-orbit coupling in silicene and two-dimensional germanium and tin},
\newblock \bibinfo{journal}{Physical Review B} \bibinfo{volume}{84}
  (\bibinfo{year}{2011}{\natexlab{b}}) \bibinfo{pages}{195430}.
\bibitem[{Xiao et~al.(2012)Xiao, Liu, Feng, Xu, and Yao}]{xiao_coupled_2012}
\bibinfo{author}{D.~Xiao}, \bibinfo{author}{G.-B. Liu},
  \bibinfo{author}{W.~Feng}, \bibinfo{author}{X.~Xu}, \bibinfo{author}{W.~Yao},
\newblock \bibinfo{title}{Coupled {Spin} and {Valley} {Physics} in {Monolayers}
  of {MoS$_2$} and {Other} {Group}-{VI} {Dichalcogenides}},
\newblock \bibinfo{journal}{Physical Review Letters} \bibinfo{volume}{108}
  (\bibinfo{year}{2012}) \bibinfo{pages}{196802}.
\bibitem[{Lu et~al.(2016)Lu, Zhou, Chang, Guan, Chen, Jiang, Jiang, Wang, Yang,
  Feng, Kawazoe, and Lin}]{lu_multiple_2016}
\bibinfo{author}{Y.~Lu}, \bibinfo{author}{D.~Zhou}, \bibinfo{author}{G.~Chang},
  \bibinfo{author}{S.~Guan}, \bibinfo{author}{W.~Chen},
  \bibinfo{author}{Y.~Jiang}, \bibinfo{author}{J.~Jiang},
  \bibinfo{author}{X.-s. Wang}, \bibinfo{author}{S.~A. Yang},
  \bibinfo{author}{Y.~P. Feng}, \bibinfo{author}{Y.~Kawazoe},
  \bibinfo{author}{H.~Lin},
\newblock \bibinfo{title}{Multiple unpinned {Dirac} points in group-{Va}
  single-layers with phosphorene structure},
\newblock \bibinfo{journal}{npj Computational Materials} \bibinfo{volume}{2}
  (\bibinfo{year}{2016}) \bibinfo{pages}{16011}.
\bibitem[{Bernevig et~al.(2006)Bernevig, Hughes, and
  Zhang}]{bernevig_quantum_2006}
\bibinfo{author}{B.~A. Bernevig}, \bibinfo{author}{T.~L. Hughes},
  \bibinfo{author}{S.-C. Zhang},
\newblock \bibinfo{title}{Quantum {Spin} {Hall} {Effect} and {Topological}
  {Phase} {Transition} in {HgTe} {Quantum} {Wells}},
\newblock \bibinfo{journal}{Science} \bibinfo{volume}{314}
  (\bibinfo{year}{2006}) \bibinfo{pages}{1757}.
\bibitem[{Wan et~al.(2011)Wan, Turner, Vishwanath, and
  Savrasov}]{wan_topological_2011}
\bibinfo{author}{X.~Wan}, \bibinfo{author}{A.~M. Turner},
  \bibinfo{author}{A.~Vishwanath}, \bibinfo{author}{S.~Y. Savrasov},
\newblock \bibinfo{title}{Topological semimetal and {Fermi}-arc surface states
  in the electronic structure of pyrochlore iridates},
\newblock \bibinfo{journal}{Phys. Rev. B} \bibinfo{volume}{83}
  (\bibinfo{year}{2011}) \bibinfo{pages}{205101}.
\bibitem[{Wang et~al.(2012)Wang, Sun, Chen, Franchini, Xu, Weng, Dai, and
  Fang}]{wang_dirac_2012}
\bibinfo{author}{Z.~Wang}, \bibinfo{author}{Y.~Sun}, \bibinfo{author}{X.-Q.
  Chen}, \bibinfo{author}{C.~Franchini}, \bibinfo{author}{G.~Xu},
  \bibinfo{author}{H.~Weng}, \bibinfo{author}{X.~Dai},
  \bibinfo{author}{Z.~Fang},
\newblock \bibinfo{title}{Dirac semimetal and topological phase transitions in
  {A}$_{\textrm{3}}${Bi} ({A}={Na}, {K}, {Rb})},
\newblock \bibinfo{journal}{Physical Review B} \bibinfo{volume}{85}
  (\bibinfo{year}{2012}) \bibinfo{pages}{195320}.
\bibitem[{Yang and Nagaosa(2014)}]{yang_classification_2014}
\bibinfo{author}{B.-J. Yang}, \bibinfo{author}{N.~Nagaosa},
\newblock \bibinfo{title}{Classification of stable three-dimensional {Dirac}
  semimetals with nontrivial topology},
\newblock \bibinfo{journal}{Nature Communications} \bibinfo{volume}{5}
  (\bibinfo{year}{2014}) \bibinfo{pages}{ncomms5898}.
\bibitem[{Young and Kane(2015)}]{young_dirac_2015}
\bibinfo{author}{S.~M. Young}, \bibinfo{author}{C.~L. Kane},
\newblock \bibinfo{title}{Dirac {Semimetals} in {Two} {Dimensions}},
\newblock \bibinfo{journal}{Physical Review Letters} \bibinfo{volume}{115}
  (\bibinfo{year}{2015}) \bibinfo{pages}{126803}.
\bibitem[{Weng et~al.(2016)Weng, Fang, Fang, and Dai}]{weng_topological_2016}
\bibinfo{author}{H.~Weng}, \bibinfo{author}{C.~Fang},
  \bibinfo{author}{Z.~Fang}, \bibinfo{author}{X.~Dai},
\newblock \bibinfo{title}{Topological semimetals with triply degenerate nodal
  points in $\theta$-phase tantalum nitride},
\newblock \bibinfo{journal}{Physical Review B} \bibinfo{volume}{93}
  (\bibinfo{year}{2016}) \bibinfo{pages}{241202}.
\bibitem[{Zhu et~al.(2016)Zhu, Winkler, Wu, Li, and
  Soluyanov}]{zhu_triple_2016}
\bibinfo{author}{Z.~Zhu}, \bibinfo{author}{G.~W. Winkler},
  \bibinfo{author}{Q.~Wu}, \bibinfo{author}{J.~Li}, \bibinfo{author}{A.~A.
  Soluyanov},
\newblock \bibinfo{title}{Triple {Point} {Topological} {Metals}},
\newblock \bibinfo{journal}{Physical Review X} \bibinfo{volume}{6}
  (\bibinfo{year}{2016}) \bibinfo{pages}{031003}.
\bibitem[{Yang et~al.(2014)Yang, Pan, and Zhang}]{PhysRevLett.113.046401}
\bibinfo{author}{S.~A. Yang}, \bibinfo{author}{H.~Pan},
  \bibinfo{author}{F.~Zhang},
\newblock \bibinfo{title}{{Dirac} and {Weyl} {Superconductors} in {Three}
  {Dimensions}},
\newblock \bibinfo{journal}{Phys. Rev. Lett.} \bibinfo{volume}{113}
  (\bibinfo{year}{2014}) \bibinfo{pages}{046401}.
\bibitem[{Weng et~al.(2015)Weng, Liang, Xu, Yu, Fang, Dai, and
  Kawazoe}]{weng_topological_2015}
\bibinfo{author}{H.~Weng}, \bibinfo{author}{Y.~Liang}, \bibinfo{author}{Q.~Xu},
  \bibinfo{author}{R.~Yu}, \bibinfo{author}{Z.~Fang}, \bibinfo{author}{X.~Dai},
  \bibinfo{author}{Y.~Kawazoe},
\newblock \bibinfo{title}{Topological node-line semimetal in three-dimensional
  graphene networks},
\newblock \bibinfo{journal}{Physical Review B} \bibinfo{volume}{92}
  (\bibinfo{year}{2015}) \bibinfo{pages}{045108}.
\bibitem[{Zhao et~al.(2016)Zhao, Schnyder, and Wang}]{zhao_unified_2016}
\bibinfo{author}{Y.~X. Zhao}, \bibinfo{author}{A.~P. Schnyder},
  \bibinfo{author}{Z.~Wang},
\newblock \bibinfo{title}{Unified {Theory} of {P} {T} and {C} {P} {Invariant}
  {Topological} {Metals} and {Nodal} {Superconductors}},
\newblock \bibinfo{journal}{Physical Review Letters} \bibinfo{volume}{116}
  (\bibinfo{year}{2016}).
\bibitem[{Bzdušek and Sigrist(2017)}]{bzdusek_robust_2017}
\bibinfo{author}{T.~Bzdušek}, \bibinfo{author}{M.~Sigrist},
\newblock \bibinfo{title}{Robust doubly charged nodal lines and nodal surfaces
  in centrosymmetric systems},
\newblock \bibinfo{journal}{Phys. Rev. B} \bibinfo{volume}{96}
  (\bibinfo{year}{2017}) \bibinfo{pages}{155105}.
\bibitem[{Wu et~al.(2018)Wu, Liu, Li, Zhong, Yu, Sheng, Zhao, and
  Yang}]{wu_nodal_2018}
\bibinfo{author}{W.~Wu}, \bibinfo{author}{Y.~Liu}, \bibinfo{author}{S.~Li},
  \bibinfo{author}{C.~Zhong}, \bibinfo{author}{Z.-M. Yu},
  \bibinfo{author}{X.-L. Sheng}, \bibinfo{author}{Y.~X. Zhao},
  \bibinfo{author}{S.~A. Yang},
\newblock \bibinfo{title}{Nodal surface semimetals: {Theory} and material
  realization},
\newblock \bibinfo{journal}{Physical Review B} \bibinfo{volume}{97}
  (\bibinfo{year}{2018}) \bibinfo{pages}{115125}.
\bibitem[{Gresch(2018)}]{gresch_identifying_2018}
\bibinfo{author}{D.~Gresch},
\newblock \bibinfo{title}{Identifying {Topological} {Semimetals}},
\newblock \bibinfo{journal}{Ph.D. thesis (ETH Zurich)}  (\bibinfo{year}{2018}).
\bibitem[{Varjas et~al.(2018)Varjas, Rosdahl, and Akhmerov}]{varjas_qsymm_2018}
\bibinfo{author}{D.~Varjas}, \bibinfo{author}{T.~. Rosdahl},
  \bibinfo{author}{A.~R. Akhmerov},
\newblock \bibinfo{title}{Qsymm: algorithmic symmetry finding and symmetric
  {Hamiltonian} generation},
\newblock \bibinfo{journal}{New Journal of Physics} \bibinfo{volume}{20}
  (\bibinfo{year}{2018}) \bibinfo{pages}{093026}.
\bibitem[{Jiang et~al.(2021)Jiang, Fang, and Fang}]{jiang_kp_2021}
\bibinfo{author}{Y.~Jiang}, \bibinfo{author}{Z.~Fang},
  \bibinfo{author}{C.~Fang},
\newblock \bibinfo{title}{A kp {Effective} {Hamiltonian} {Generator}},
\newblock \bibinfo{journal}{Chin. Phys. Lett.} \bibinfo{volume}{38}
  (\bibinfo{year}{2021}) \bibinfo{pages}{077104}.
\bibitem[{Zhan et~al.(2021)Zhan, Shi, Yang, and Zhang}]{zhan_programmable_2021}
\bibinfo{author}{G.~Zhan}, \bibinfo{author}{M.~Shi}, \bibinfo{author}{Z.~Yang},
  \bibinfo{author}{H.~Zhang},
\newblock \bibinfo{title}{A {Programmable} k \${\textbackslash}cdotp\$ p
  {Hamiltonian} {Method} and {Application} to {Magnetic} {Topological}
  {Insulator} {MnBi2Te4}},
\newblock \bibinfo{journal}{Chinese Physics Letters} \bibinfo{volume}{38}
  (\bibinfo{year}{2021}) \bibinfo{pages}{077105}.
\bibitem[{Luks et~al.(1997)Luks, R\'{a}k\'{o}czi, and
  Wright}]{luks_algorithms_1997}
\bibinfo{author}{E.~M. Luks}, \bibinfo{author}{F.~R\'{a}k\'{o}czi},
  \bibinfo{author}{C.~R. Wright},
\newblock \bibinfo{title}{Some {Algorithms} for {Nilpotent} {Permutation}
  {Groups}},
\newblock \bibinfo{journal}{Journal of Symbolic Computation}
  \bibinfo{volume}{23} (\bibinfo{year}{1997}) \bibinfo{pages}{335--354}.
\bibitem[{Zhang et~al.(2017)Zhang, Yu, Sheng, Yang, and
  Yang}]{zhang_coexistence_2017}
\bibinfo{author}{X.~Zhang}, \bibinfo{author}{Z.-M. Yu}, \bibinfo{author}{X.-L.
  Sheng}, \bibinfo{author}{H.~Y. Yang}, \bibinfo{author}{S.~A. Yang},
\newblock \bibinfo{title}{Coexistence of four-band nodal rings and triply
  degenerate nodal points in centrosymmetric metal diborides},
\newblock \bibinfo{journal}{Physical Review B} \bibinfo{volume}{95}
  (\bibinfo{year}{2017}) \bibinfo{pages}{235116}.
\bibitem[{Yu et~al.(2022)Yu, Zhang, Liu, Wu, Li, Zhang, Yang, and
  Yao}]{yu_encyclopedia_2021}
\bibinfo{author}{Z.-M. Yu}, \bibinfo{author}{Z.~Zhang}, \bibinfo{author}{G.-B.
  Liu}, \bibinfo{author}{W.~Wu}, \bibinfo{author}{X.-P. Li},
  \bibinfo{author}{R.-W. Zhang}, \bibinfo{author}{S.~A. Yang},
  \bibinfo{author}{Y.~Yao},
\newblock \bibinfo{title}{Encyclopedia of emergent particles in
  three-dimensional crystals},
\newblock \bibinfo{journal}{Science Bulletin} \bibinfo{volume}{67}
  (\bibinfo{year}{2022}) \bibinfo{pages}{375--380}.
\bibitem[{Liu et~al.(2022)Liu, Zhang, Yu, Yang, and Yao}]{liu_systematic_2022}
\bibinfo{author}{G.-B. Liu}, \bibinfo{author}{Z.~Zhang}, \bibinfo{author}{Z.-M.
  Yu}, \bibinfo{author}{S.~A. Yang}, \bibinfo{author}{Y.~Yao},
\newblock \bibinfo{title}{Systematic investigation of emergent particles in
  type-{III} magnetic space groups},
\newblock \bibinfo{journal}{Physical Review B} \bibinfo{volume}{105}
  (\bibinfo{year}{2022}) \bibinfo{pages}{085117}.
\bibitem[{Zhang et~al.(2022)Zhang, Liu, Yu, Yang, and
  Yao}]{zhang_encyclopedia_2022}
\bibinfo{author}{Z.~Zhang}, \bibinfo{author}{G.-B. Liu}, \bibinfo{author}{Z.-M.
  Yu}, \bibinfo{author}{S.~A. Yang}, \bibinfo{author}{Y.~Yao},
\newblock \bibinfo{title}{Encyclopedia of emergent particles in type-{IV}
  magnetic space groups},
\newblock \bibinfo{journal}{Physical Review B} \bibinfo{volume}{105}
  (\bibinfo{year}{2022}) \bibinfo{pages}{104426}.
\bibitem[{Hogben(2006)}]{hogben_handbook_2006}
\bibinfo{editor}{L.~Hogben} (Ed.), \bibinfo{title}{Handbook of {Linear}
  {Algebra}}, \bibinfo{year}{2006}.
\bibitem[{Zhang et~al.(2022)Zhang, Yu, Liu, and Yao}]{zhang_magnetictb_2021}
\bibinfo{author}{Z.~Zhang}, \bibinfo{author}{Z.-M. Yu}, \bibinfo{author}{G.-B.
  Liu}, \bibinfo{author}{Y.~Yao},
\newblock \bibinfo{title}{{MagneticTB}: {A} package for tight-binding model of
  magnetic and non-magnetic materials},
\newblock \bibinfo{journal}{Computer Physics Communications}
  \bibinfo{volume}{270} (\bibinfo{year}{2022}) \bibinfo{pages}{108153}.
\bibitem[{Herring(1937)}]{herring_effect_1937}
\bibinfo{author}{C.~Herring},
\newblock \bibinfo{title}{Effect of {Time}-{Reversal} {Symmetry} on {Energy}
  {Bands} of {Crystals}},
\newblock \bibinfo{journal}{Physical Review} \bibinfo{volume}{52}
  (\bibinfo{year}{1937}) \bibinfo{pages}{361}.
\bibitem[{Bradley and Cracknell(2009)}]{bradley_mathematical_2009}
\bibinfo{author}{C.~Bradley}, \bibinfo{author}{A.~Cracknell},
  \bibinfo{title}{Mathematical theory of symmetry in solids: representation
  theory for point groups and space groups}, Oxford classic texts in the
  physical sciences, \bibinfo{address}{Oxford}, \bibinfo{year}{2009}.
\bibitem[{Bradley and Davies(1968)}]{bradley_magnetic_1968}
\bibinfo{author}{C.~J. Bradley}, \bibinfo{author}{B.~L. Davies},
\newblock \bibinfo{title}{Magnetic {Groups} and {Their} {Corepresentations}},
\newblock \bibinfo{journal}{Reviews of Modern Physics} \bibinfo{volume}{40}
  (\bibinfo{year}{1968}) \bibinfo{pages}{359--379}.
\bibitem[{Lax(2001)}]{lax_symmetry_2001}
\bibinfo{author}{M.~Lax}, \bibinfo{title}{Symmetry {Principles} in {Solid}
  {State} and {Molecular} {Physics}}, \bibinfo{year}{2001}.
\bibitem[{Dresselhaus et~al.(2008)Dresselhaus, Dresselhaus, and
  Jorio}]{dresselhaus_group_2008}
\bibinfo{author}{M.~S. Dresselhaus}, \bibinfo{author}{G.~Dresselhaus},
  \bibinfo{author}{A.~Jorio}, \bibinfo{title}{Group theory: application to the
  physics of condensed matter}, \bibinfo{address}{Berlin},
  \bibinfo{year}{2008}.
\bibitem[{Aroyo et~al.(2006)Aroyo, Kirov, Capillas, Perez-Mato, and
  Wondratschek}]{aroyo_bilbao_2006}
\bibinfo{author}{M.~I. Aroyo}, \bibinfo{author}{A.~Kirov},
  \bibinfo{author}{C.~Capillas}, \bibinfo{author}{J.~M. Perez-Mato},
  \bibinfo{author}{H.~Wondratschek},
\newblock \bibinfo{title}{Bilbao {Crystallographic} {Server}. {II}.
  {Representations} of crystallographic point groups and space groups},
\newblock \bibinfo{journal}{Acta Crystallographica Section A}
  \bibinfo{volume}{62} (\bibinfo{year}{2006}) \bibinfo{pages}{115--128}.
\bibitem[{Bosma et~al.(1997)Bosma, Cannon, and Playoust}]{bosma_magma_1997}
\bibinfo{author}{W.~Bosma}, \bibinfo{author}{J.~Cannon},
  \bibinfo{author}{C.~Playoust},
\newblock \bibinfo{title}{The {Magma} {Algebra} {System} {I}: {The} {User}
  {Language}},
\newblock \bibinfo{journal}{Journal of Symbolic Computation}
  \bibinfo{volume}{24} (\bibinfo{year}{1997}) \bibinfo{pages}{235--265}.
\bibitem[{Iraola et~al.(2020)Iraola, Ma\~{n}es, Bradlyn, Neupert, Vergniory,
  and Tsirkin}]{iraola_irrep_2020}
\bibinfo{author}{M.~Iraola}, \bibinfo{author}{J.~L. Ma\~{n}es},
  \bibinfo{author}{B.~Bradlyn}, \bibinfo{author}{T.~Neupert},
  \bibinfo{author}{M.~G. Vergniory}, \bibinfo{author}{S.~S. Tsirkin},
\newblock \bibinfo{title}{{IrRep}: symmetry eigenvalues and irreducible
  representations of ab initio band structures},
\newblock \bibinfo{journal}{arXiv:2009.01764 [cond-mat, physics:physics]}
  (\bibinfo{year}{2020}).
\bibitem[{Matsugatani et~al.(2021)Matsugatani, Ono, Nomura, and
  Watanabe}]{matsugatani_qeirreps_2021}
\bibinfo{author}{A.~Matsugatani}, \bibinfo{author}{S.~Ono},
  \bibinfo{author}{Y.~Nomura}, \bibinfo{author}{H.~Watanabe},
\newblock \bibinfo{title}{qeirreps: {An} open-source program for {Quantum}
  {ESPRESSO} to compute irreducible representations of {Bloch} wavefunctions},
\newblock \bibinfo{journal}{Computer Physics Communications}
  \bibinfo{volume}{264} (\bibinfo{year}{2021}) \bibinfo{pages}{107948}.
\bibitem[{noa(2021)}]{noauthor_gap_2021}
\bibinfo{title}{{GAP} -- {Groups}, {Algorithms}, and {Programming}, {Version}
  4.11.1}, \bibinfo{year}{2021}.
\bibitem[{Liu et~al.(2021)Liu, Chu, Zhang, Yu, and
  Yao}]{liu_spacegroupirep_2021}
\bibinfo{author}{G.-B. Liu}, \bibinfo{author}{M.~Chu},
  \bibinfo{author}{Z.~Zhang}, \bibinfo{author}{Z.-M. Yu},
  \bibinfo{author}{Y.~Yao},
\newblock \bibinfo{title}{{SpaceGroupIrep}: {A} package for irreducible
  representations of space group},
\newblock \bibinfo{journal}{Computer Physics Communications}
  \bibinfo{volume}{265} (\bibinfo{year}{2021}) \bibinfo{pages}{107993}.
\bibitem[{Liu and et~al.(2021)}]{liu_MSGCorep_2021}
\bibinfo{author}{G.-B. Liu}, \bibinfo{author}{et~al.},
\newblock \bibinfo{title}{{MSGCorep}: {A} package for corepresentations of
  magnetic space groups, to be published.}  (\bibinfo{year}{2021}).
\bibitem[{Cormen(2001)}]{cormen_introduction_2001}
\bibinfo{editor}{T.~H. Cormen} (Ed.), \bibinfo{title}{Introduction to
  algorithms}, \bibinfo{address}{Cambridge, Mass}, \bibinfo{year}{2001}.
\bibitem[{Tang and Wan(2021)}]{tang_exhaustive_2021}
\bibinfo{author}{F.~Tang}, \bibinfo{author}{X.~Wan},
\newblock \bibinfo{title}{Exhaustive constructions of effective models in 1651
  magnetic space groups},
\newblock \bibinfo{journal}{Physical Review B} \bibinfo{volume}{104}
  (\bibinfo{year}{2021}) \bibinfo{pages}{085137}.

\end{thebibliography}

\end{document}